\documentclass{elsart}
\input{epsf}

\begin{document}

\begin{frontmatter}

\title{Singular Character of Critical Points in Nuclei}

\author{V. Werner$^{1)}$, P. von Brentano$^{1)}$, R.F. Casten$^{2)}$,
  and J. Jolie$^{1)}$}

\address{$^{1)}$Institut f\"{u}r Kernphysik, Universit\"at zu
K\"oln, Germany\\ $^{2)}$WNSL, Yale University, New Haven, CT
06520-8124, USA}

\begin{abstract}
The concept of critical points in nuclear phase transitional
regions is discussed from the standpoints of Q-invariants, simple
observables and wave function entropy. It is shown that these
critical points very closely coincide with the turning points of
the discussed quantities, establishing the singular
character of these points in nuclear phase transition regions between
vibrational and rotational nuclei, with a finite number of particles.
\end{abstract}


\maketitle
\end{frontmatter}

Nuclear structural evolution in transitional regions is often
thought of as a continuous variation of properties, as a function
of nucleon number, from one idealized limit (e.g., vibrator,
rotor) to another.  The rapidity of structural change may vary
across a transitional sequence of nuclei, and different mass
regions exhibit different rates of change but, until recently, no
individual point along these evolutionary trajectories could be
singled out with special observational properties.

In the last years, however, the concept of critical points in
shape/phase transition regions has been much discussed
\cite{Wol94,Cas93,Iac98,Cas99,JJ99}. 
While the concept itself is well known in nuclei (in the context
of the coherent state formalism \cite{Die80,Gin80} of the IBA model
\cite{IacAri}),
it is only very recently that analytic descriptions of critical point
nuclei have been given \cite{Iac01,Iac00}. This is a significant point since,
historically, such  nuclei have been the most difficult to treat:
they exhibit competing degrees of freedom, and one has had to
resort to numerical calculations.

Two critical point symmetries, called E(5) and X(5), have been proposed
\cite{Iac01,Iac00}, giving analytic expressions for observables which
are exactly at the critical points of a vibrator to axially asymmetric 
($\gamma$-soft) rotor transition region, and of a vibrator to symmetric rotor
transition region, respectively, for an infinite number of nucleons. An
important aspect of this is that, for the first time, one is able to
associate special observational characteristics to a specific
point along a trajectory from one structural limit to another.
Recently \cite{JJ01}, using the methods presented here,
the well-known O(6) limit of the IBA has also been
identified as another, heretofore unrecognized, critical point
symmetry, for the transition between prolate and oblate nuclei. This is an
important result since the O(6) symmetry can be calculated in the IBA
for finite nucleon numbers, in contrast to the non-IBA symmetries E(5)
and X(5). So far only two examples for nuclei \cite{Cas01,Cas00} which
lie close to the X(5) and E(5) symmetries are known while,
interestingly, there are many examples for O(6) like nuclei. 
In the present work we will restrict our discussion to prolate nuclei.

To understand the evolution of structure in real nuclei, with a finite
number of nucleons, it is important to gather information about
systematic changes of observables at or near such critical
points. This aim can be achieved by the use of a model that is able to
describe limiting cases of nuclear structure - vibrators,
rotors and $\gamma$-soft nuclei - and a large variety of nuclei
between these limits. Such a model is given by the IBA, which - in the 
expansion of the coherent state formalism - exhibits critical
points as has been discussed in refs. \cite{Die80,Gin80,Lop96,Lop98}. 
We stress that
the critical point descriptions X(5) and E(5) are defined in terms of
a geometrical approach, not the IBA. Nevertheless the IBA provides a
convenient tool to span a range of structure, including phase
transitions, and also to assess effects of finite particle numbers.

It is the purpose of this Letter to show, from several
complementary theoretical approaches, that there is independent
evidence for the singular character of these critical points, and
independent ways of identifying them in observables calculated in
collective models. To do so we bring together three
major themes: the already mentioned study of phase transitional
regions and critical point nuclei, the behavior of quadrupole shape
(Q)-invariants, and the study of chaos and entropy in nuclear systems.
We show that the
critical points occur very near to the turning points (points of
steepest descent or ascent) of these Q-invariants -- that is, at the
extrema of their first derivatives. The same behavior will also
be shown to hold for some more easily accessible observables.

To span the transition regions, it is
convenient to use the IBA Hamiltonian in the following form

\begin{equation}
\label{eq:hecqf}
H=a[(1-\zeta)n_d-\frac{\zeta}{4N}Q\cdot{Q}]
\end{equation}

\noindent where
$Q=s^\dag\widetilde{d}+d^\dag{s}+\chi[d^\dag\widetilde{d}]^{(2)}$ and
we consider the well known parameter space of the extended consistent
Q formalism (ECQF) varying $\zeta$ between 0 and 1, and $\chi$ from 0
to $-\sqrt{7}/2=-1.32$, while $a$ is a scaling factor. This
parametrization is equivalent to the more commonly encountered
(equivalent) ECQF \cite{War82,Lip85} form of $H$, which includes the
parameters $\epsilon$ and $\kappa$.

\begin{figure}[ht]
\epsfxsize 6.5cm
\epsfbox{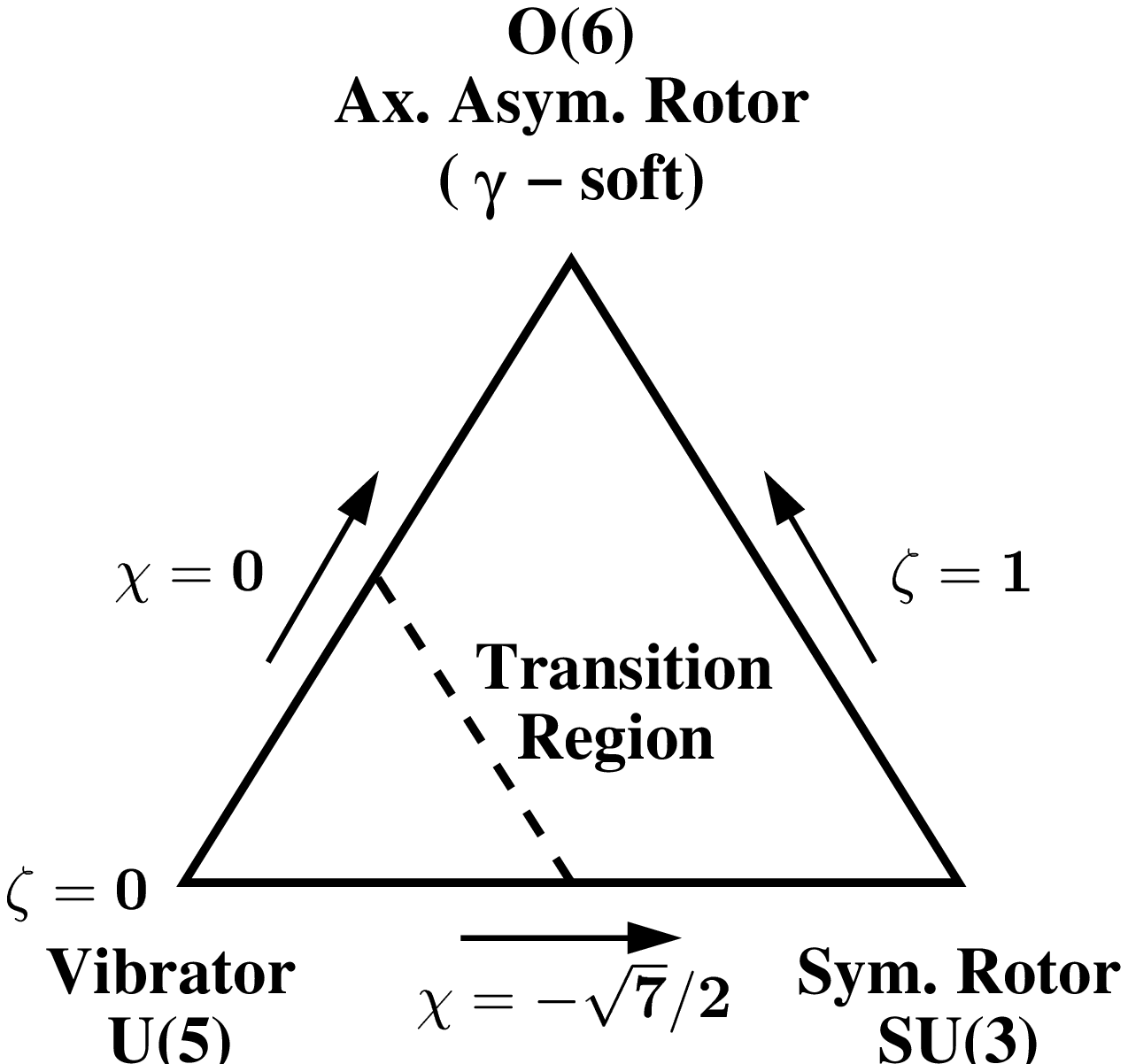}
\caption{Symmetry triangle of the IBA model. The
U(5)$\leftrightarrow$O(6) leg is characterized by $\chi=0$ and varying 
$\zeta$, while the U(5)$\leftrightarrow$SU(3) transition region has
$\chi=-\sqrt{7}/2$ and $\zeta$ is varied. The dashed line indicates 
the phase transitional region where critical points are found.}
\label{fig:triangle} 
\end{figure}

Figure \ref{fig:triangle} illustrates the three dynamical symmetries
of the IBA in terms of a triangle. With the Hamiltonian of
Eq. (\ref{eq:hecqf}) it is easy to calculate the structure for any
point in the triangle.
For $\zeta=0$ one obtains a U(5) structure (for any $\chi$), and
$\zeta=1,\chi=-\sqrt{7}/2$ gives SU(3).  Thus, a
U(5)$\leftrightarrow$SU(3) transition region is defined by
$\chi=-\sqrt{7}/2$ and $\zeta$ varying from 0 to 1, while a
U(5)$\leftrightarrow$O(6) region has $\chi=0$ and $\zeta$ varying
from 0 to 1.

One can use the coherent state formalism \cite{Die80,Gin80} of the IBA
model to identify the critical points in the ECQF space. In this
approach, the energy functional for the ECQF Hamiltonian is given by
\scriptsize
\begin{equation}
\label{eq:ecsf}
{{E}(\zeta,\chi,\beta,\gamma)=\frac{N\beta^2(1-\frac{\zeta(\chi^2-3)}{4N-4N\zeta+\zeta})}{1+\beta^2}-\frac{\frac{N(N-1)\zeta}{4N-4N\zeta+\zeta}\left(4\beta^2-4\sqrt{\frac{2}{7}}\chi\beta^3\cos{3\gamma}+\frac{2}{7}\chi^2\beta^4\right)}{(1+\beta^2)^2}}
\end{equation}
\normalsize
The variation of $\zeta$ changes the structure between the
vibrator limit and rotational nuclei -- both axially symmetric and
axially asymmetric -- which are the transitions we will focus on.
Critical points in $\zeta$ are found where E becomes flat at
$\beta=0$. These points, which we refer to as $\zeta_c$, can be
derived by evaluating the condition
\begin{equation}
\label{eq:critcond}
\left|\frac{\partial^2{E(\zeta_c)}}{\partial\beta^2}\right|_{\beta=0}=0 
\ .
\end{equation}
On the transition path from U(5) to O(6) (for $\chi=0$) exactly one
critical point is found, namely where a second, deformed, minimum in
$\beta$ of the energy functional emerges. 

\begin{figure}[ht]
\epsfxsize 7cm
\epsfbox{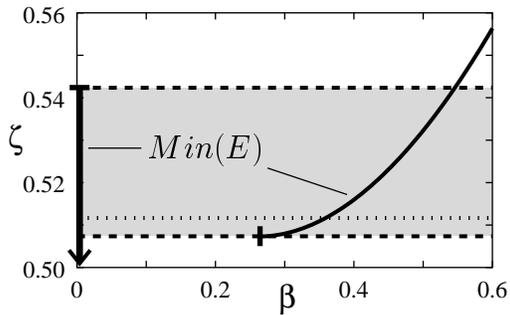}
\caption{The thick lines represent the locus in the ($\zeta$,$\beta$)
parameter space where the energy functional of the coherent state
formalism has a local minimum. The thick line at $\beta=0$ extends
downwards to $\zeta=0$. The results are shown for the case of
N=10 bosons. Dashed lines mark critical $\zeta$ values where one
minimum disappears (the spherical one at and above the larger value,
the deformed one at and below the lower value). Only in the shaded
area two minima coexist. The dotted line marks the critical $\zeta$
value where these two minima are equally deep.}
\label{fig:critreg} 
\end{figure}

The situation becomes more
complicated for transitions with $\chi\ne 0$. In these cases, the
spherical minimum is joined by a deformed minimum and both minima
coexist in a very close parameter range in $\zeta$, converging to one
point when approaching $\chi=0$. 
Thus, in general there exist three critical points, which is
illustrated in Fig. \ref{fig:critreg} for N=10 bosons for the limiting
case of $\chi=-\sqrt{7}/2$. The thick lines in Fig. \ref{fig:critreg} give
points in the $(\zeta,\beta)$ plane, which are local minima of the
energy functional (\ref{eq:ecsf}). The shaded area is the parameter
range of $\zeta$, where two local minima of
the energy functional coexist. The lower dashed line gives the
critical $\zeta$ value where a deformed minimum appears, while the
upper dashed line gives the critical point in $\zeta$ where the
spherical minimum disappears and only the deformed minimum is left.
The dotted line gives the critical $\zeta$ value where two coexisting
minima are equally deep. The parameter region in between is small for
any boson number.

Thus, as it is the aim of this work to identify the critical points in
observables, and we do not expect to be able to distinguish between these
three points (close lying in $\zeta$) in real nuclei, we restrict
ourselves to the critical point given by condition
(\ref{eq:critcond}) where the spherical minimum disappears, and which
is given by
\begin{equation}
\label{eq:crittrans}
\zeta_c = \frac{4N}{8N-8+\chi^2} \
\stackrel{N\rightarrow\infty}{\longrightarrow} \ 0.5 \ .
\end{equation}
The $\chi$ dependence is just a finite N effect, and thus it is
convenient to vary only the parameter $\zeta$ for the investigation of 
phase transitions between vibrational and rotational
nuclei. Additionally we note that the choice of our 
parametrization has the convenient feature that in the large N limit
we get $\zeta_c=0.5$.

While, due to their physical meaning, the endpoints of the line of
critical points between $\chi=0$ and $\chi=-\sqrt{7}/2$ in
Fig. \ref{fig:triangle} can be approximately related to the non-IBA
symmetries E(5) and X(5), we see that a much richer structure shows up
in the IBA, where critical points occur over the whole transitional
region between these legs of the symmetry triangle.

Since we are interested in obtaining signatures for critical points in 
observables including matrix elements, we now survey the behavior of
Q-invariants \cite{Cli86,Kum72} in the transition regions.
Recently, the concept of Q-invariants has been
re-investigated in the framework of the IBA model and the Q-phonon
approach \cite{Sie94,Ots94}, and the behavior of these moments across the
gamut of nuclear collective structures has been elucidated
\cite{Jol97,Pal98,Wer00}. These invariants represent quadratic and higher order
moments of the quadrupole operator. The invariants are denoted
$q_n$ and $K_n\equiv{q_n/q_2^{n/2}}$, and are defined by
expressions of the generic type

\begin{equation}
\label{eq:qngen}
q_n\sim\langle\Psi_0|Q_1\cdot{Q_2}\cdots{Q_n}|\Psi_0\rangle
\end{equation}

where $\Psi_0$ is the ground state wave function, and where
intermediate angular momentum couplings in the operator are
omitted for simplicity.

For the IBA \cite{IacAri}, the Q-invariants have been evaluated over the
entire symmetry triangle of Fig. \ref{fig:triangle}. To show the
extreme cases, we first focus on the two
transition paths U(5)$\leftrightarrow$SU(3) ($\chi=-\sqrt{7}/2$) and
U(5)$\leftrightarrow$O(6) ($\chi=0$). We note that the
invariants $q_2$, $K_3$, $K_4$, and $\sigma_\gamma\equiv{K_6-K_3^2}$
represent, respectively, the quadrupole deformation, the triaxiality, 
the softness of the nuclear shape in $\beta$, and in $\gamma$.

\begin{figure}[ht]
\epsfxsize 14cm
\epsfbox{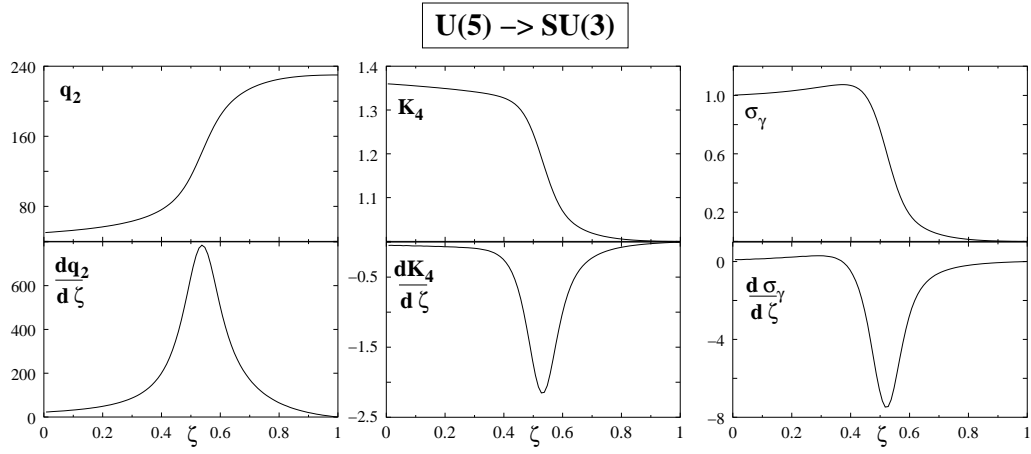}
\caption{Behavior of $q_2$, $K_4$ and $\sigma_\gamma$, and
their first derivatives with respect to $\zeta$, for the
U(5)$\leftrightarrow$SU(3) transition region, calculated for N=10 bosons.}
\label{fig:ku5su3} 
\end{figure}

We first study the U(5)$\leftrightarrow$SU(3) transition and
obtain the results shown for N=10 in the top row of
Fig. \ref{fig:ku5su3} for $q_2,K_4$ and $\sigma_{\gamma}$.  Each of
these exhibits a rapidly changing behavior which has a turning point
$\zeta_t$ near $\zeta=0.5$. To investigate this in more detail, the
second row of Fig. \ref{fig:ku5su3} shows the first derivatives with
respect to $\zeta$.  Again there is a striking consistency of
behavior: the first derivative has an extremum at essentially the same
point for each invariant.

Specifically, the turning points (the zeros of the second
derivatives) are: $\zeta_t=0.54$ for $q_2$; $\zeta_t=0.53$ for
$K_4$; and $\zeta_t=0.52$ for $\sigma_\gamma$.  In the coherent
state formalism, for N=10, one obtains $\zeta_c=0.54$ for the
U(5)$\leftrightarrow$SU(3) case. This is very close to the turning
points in $q_2, K_4$ and $\sigma_\gamma$: that is
$\zeta_t\sim\zeta_c$. This correspondence between the turning
points and the critical points is the main result of this work.
The small differences probably represent a finite boson number
effect.

\begin{figure}[ht]
\epsfxsize 8.5cm
\epsfbox{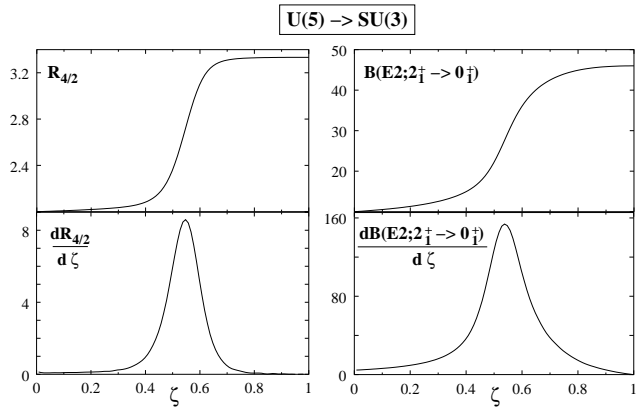}
\caption{Similar to Fig. 3 (for N=10) for the observables $R_{4/2}$ and
$B(E2:2^+_1\rightarrow{0^+_1})$ for the U(5)$\leftrightarrow$SU(3)
transition region.}
\label{fig:obsu5su3} 
\end{figure}

This identification of a special point along the structural
evolution from vibrator to rotor is apparent even in the simplest
observables as well. In Fig. \ref{fig:obsu5su3} we show the behavior of the
structural observables $R_{4/2}\equiv{E}(4^+_1)/E(2^+_1)$ and
$B(E2:2^+_1\rightarrow{0^+_1)}$ for the U(5)$\leftrightarrow$SU(3)
transition, again for N=10. Clearly, as seen in the first
derivative plots in the second row, both quantities exhibit their
steepest rates of change near the critical points.  Here, the
first derivative has an extremum at $\zeta_t=0.54$ for both
$R_{4/2}$ and the B(E2) value. In this latter case, this result is
not surprising since this B(E2) value and $q_2$ are directly
related.

The existence of three critical points on the
U(5)$\leftrightarrow$SU(3) transition path seems
not to be reflected in the Q-invariants, which may be explained by the
very compact parameter region in $\zeta$ where these critical points
occur, while the peaks in the derivatives have a certain width. 
Also note that fluctuations, resulting from the limited numerical
accuracy of the {\sc Phint} code used for these calculations,
have been smoothed by the use of splines. Thus, perhaps the three
critical points just cannot be resolved in the observables due to
numerical truncations.

\begin{figure}[ht]
\epsfxsize 14cm
\epsfbox{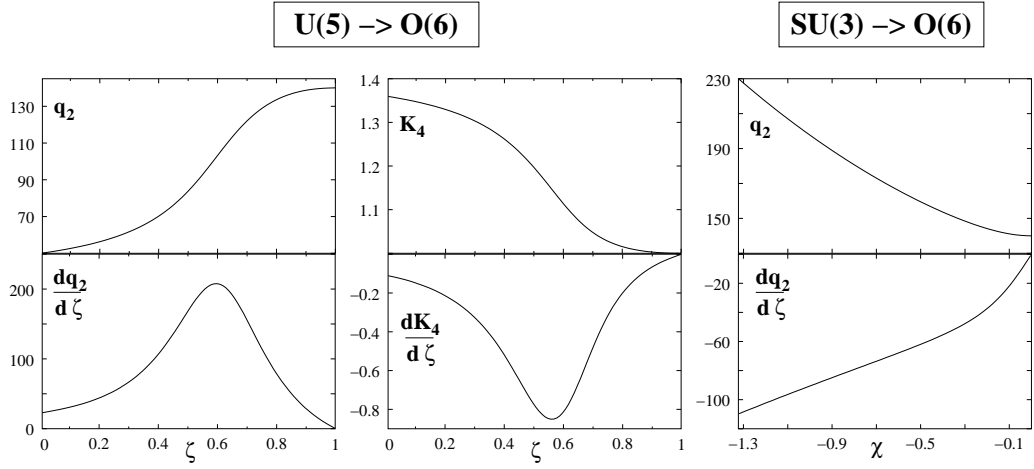}
\caption{Similar to Fig. 3 (for N=10), for $q_2$ and $K_4$, for the
U(5)$\leftrightarrow$O(6) transition region (left panels), and for
$q_2$ in the O(6)$\leftrightarrow$SU(3) (note here with respect to
$\chi$) transition region (right).}
\label{fig:ku5o6} 
\end{figure}

Returning to the Q-invariants, similar results apply in the
U(5)$\rightarrow$O(6) region. Fig. \ref{fig:ku5o6} (left panels) shows
this for $q_2$ and $K_4$. In this case the turning points (determined from
the rates of change), are: $\zeta_t=0.60$ for $q_2$ and
$\zeta_t=0.56$ for $K_4$. From Eq. (\ref{eq:ecsf}), the coherent state
formalism gives $\zeta_c=0.56$ for N=10. Again the $\zeta_t$ and
$\zeta_c$ values obtained from the behavior of the Q-invariants and
from the coherent state formalism are quite close. Lastly, we note
that the rate of change of $q_2$ and $K_4$ in the
U(5)$\leftrightarrow$O(6) case is much less than in the first order
U(5)$\rightarrow$SU(3) transition region. For example
$(dq_2/d\zeta)_{max}\sim800$ for U(5)$\leftrightarrow$SU(3) while it
is only $\sim$200 for U(5)$\leftrightarrow$O(6).  Also, the widths of
the first derivative curves are much wider (corresponding to a more
gradual structural evolution) in the U(5)$\leftrightarrow$O(6) case.

\begin{figure}[ht]
\epsfxsize 7cm
\epsfbox{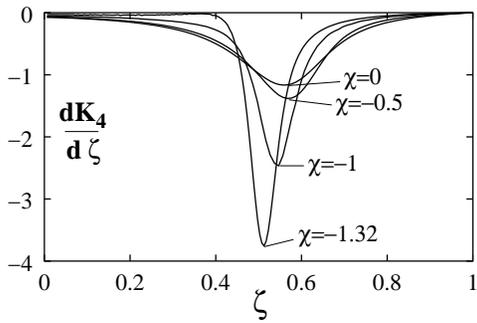}
\caption{First derivative of $K_4$ (for N=10), but for various values of the
parameter $\chi$. A peak indicating a phase transition occurs for
every value of $\chi$. The dependence of its position on $\chi$ is a
finite N effect.}
\label{fig:dk4path} 
\end{figure}

Using the IBA, it is also possible to investigate internal paths in
the symmetry triangle. In particular internal straight line
trajectories, starting from U(5), will correspond to $\chi$ values
between 0 and $-\sqrt{7}/2$, allowing a full mapping of transitional
trajectories. We illustrate such results by showing the
change of the first derivative of the shape invariant $K_4$ for 
various values of $\chi$ in Fig. \ref{fig:dk4path}. The
minima of the derivatives follow the line of critical points that is
also given in the coherent state formalism, with only a small $\chi$
dependence.

Finally, in regard to Q-invariants, we look at the
O(6)$\leftrightarrow$SU(3) transitional region.  The right panels
of Fig. \ref{fig:ku5o6} show the behavior of $q_2$ and its
derivative. Note that the shape is qualitatively different than in the
other transition regions, showing a gradually asymptotic curve and a
first derivative against $\chi$ (the appropriate variable for this
region) which is monotonic.  No critical point is definable in
this region of $\chi$ values, except when O(6) itself is reached (see
ref. \cite{JJ01}).

Another theme in nuclear structure recently has been the study of
order and chaos for different structures. It was shown in ref.
\cite{Alh91} that nuclear systems display ordered spectra at and near the
three symmetry limits of the IBA, but that there is a rapid onset
of chaotic behavior away from these benchmark regions. (See Fig.
1 of ref. \cite{Alh91} but note that the symmetry triangle is differently
defined therein.) Recently, Cejnar and Jolie \cite{Cej98a,Cej98b} have
developed the concept of wave function entropy as an alternate
(and physically intuitive) way of studying the relative complexity
of nuclear wave functions. Basically, the entropy of a state is a
measure of its spreading within a given basis. Note that this is
not the same as the chaoticity (which is basis invariant) since a
wave function may have high entropy in one basis [e.g., U(5)] and
low entropy in another [e.g., SU(3)].

Now that we showed a visible effect of critical points in various 
observables, it is interesting to see whether effects of a phase
transition can also be seen in the wave functions and thus the wave
function entropy. A rise of the wave function entropy can be expected in
moving from one limit to another, but the question is whether it also
appears in a close region with turning points which coincide with the
turning points of the previously mentioned observables. Thus, we define
\cite{Cej98a} a quantity, called $W^{\mathcal{B}}_\Psi$, for a 
state $\Psi$, that can be written in the basis ${\mathcal{B}}$ as
$\Psi=\sum^n_{i_{\mathcal{B}}}a_{i_{\mathcal{B}}}|\Psi_{\mathcal{B}}>$,
as

\begin{equation}
\label{eq:entropy}
W^{{\mathcal{B}}}_\Psi\equiv -\sum^n_{i_{{\mathcal{B}}}=1}|a_{i_{{\mathcal{B}}}}|^2\ln{|a^{{\mathcal{B}}}_{i_{{\mathcal{B}}}}|^2}
\end{equation}

\noindent where $n$ is the number of basis vectors.  If $\Psi$
coincides with a basis vector, then $W^{\mathcal{B}}_\Psi=0$.  If
$\Psi$ is uniformly spread out over the basis ${\mathcal{B}}$,
then $W^{\mathcal{B}}_\Psi\approx \ln{n}$.

A physically intuitive expression of the entropy is the quantity
\cite{Cej98b}

\begin{equation}
\label{eq:neff}
n^{{\mathcal{B}}}_{{eff}_\Psi}\equiv{exp}\ W^{\mathcal{B}}_\Psi
\end{equation}

\noindent which expresses a kind of "effective number" of wave
function components.  For a "pure" state $\Psi$,
$n^{\mathcal{B}}_{{eff}_\Psi}=1$ and for a fully de-localized
state $n^{\mathcal{B}}_{{eff}_\Psi}\approx n$.

To properly normalize the entropies we define the entropy ratio

\begin{equation}
\label{eq:rb}
r^{\mathcal{B}}\equiv\frac{{exp}\
W^{\mathcal{B}}_\Psi-1}{exp\langle W_{GOE}\rangle -1}
\end{equation}

\noindent relative to that for the Gaussian Orthogonal Ensemble \cite{Cej98b}.
The ratio $r^{\mathcal{B}}_\Psi$ varies from 0 for a pure
(localized in the basis ${\mathcal{B}}$) state to $\sim$1 for a
highly mixed state (see ref. \cite{Cej98b} for a more detailed discussion
of this normalization).

\begin{figure}[ht]
\epsfxsize 8.5cm
\epsfbox{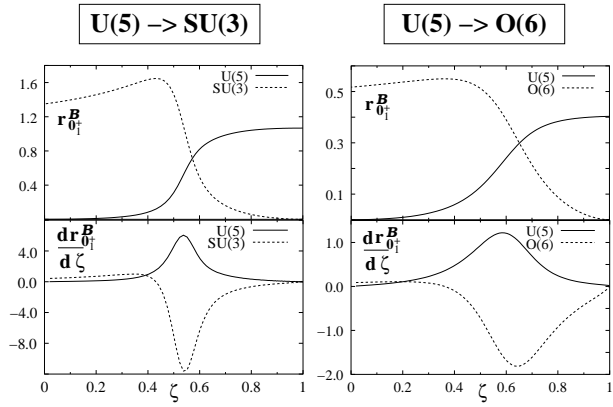}
\caption{The entropy ratio (for N=10) for the $0^+_1$ state
(top row) in the three transition regions, plotted against $\zeta$
and given, for each region, in two bases
as indicated [e.g., U(5) and SU(3) for the
U(5)$\leftrightarrow$SU(3) transition].  The lower panels give the
derivative of the entropy ratio against $\zeta$ in the
appropriate basis.}
\label{fig:entropy} 
\end{figure}

We show the results in Fig. \ref{fig:entropy} for
$r^{\mathcal{B}}_{0^+_1}$ and its derivative as a function of the
order parameter $\zeta$ for the U(5)$\leftrightarrow$SU(3) and
U(5)$\leftrightarrow$O(6)
transition regions (all for N=10). The entropy ratio for the ground
state undergoes a very rapid change near $\zeta_c$ for both
transition regions. We note that for larger boson numbers N the
transition becomes much sharper (see Fig. 6 in \cite{JJ01}). For the
U(5)$\leftrightarrow$SU(3) and U(5)$\leftrightarrow$O(6) phase
transitions, it is easy to read the turning points, $\zeta_t$,
values from the derivative plots, obtaining $\zeta_t=0.52$ and
$\zeta_t=0.59$ [in a U(5) basis], respectively, compared to values
of $\zeta_c=0.54$ and $\zeta_c=0.56$ from the coherent state
formalism. We note that the steepness of the entropy functions against
$\zeta$ increases with boson number N, as pointed out in
ref. \cite{Cej00}. This also holds true for the observables studied
above.

To conclude, from the behavior of several rather different
quantities, the Q-invariants, the simple observables $R_{4/2}$ and
$B(E2:2^+_1\rightarrow{0^+_1})$, and the wave function entropy, we
have shown that critical points of the phase transitional regions
U(5)$\leftrightarrow$SU(3) and U(5)$\leftrightarrow$O(6) are reflected 
in the behavior of these observables along these evolutionary
trajectories. This result was obtained for finite boson numbers,
making it possible to investigate effects of valence particle number
on the singularities.

We are grateful to N.V. Zamfir, F. Iachello, J. Eberth and K.
Heyde for useful discussions, and to P. Cejnar for the entropy
calculations. Work supported by the U.S.DOE under Grant number
DE-FG02-91ER40609 and by the DFG under Project number Br 799/10-1
and by NATO Research Grant no. 950668. One of us [RFC] is grateful
to the Institut f\"{u}r Kernphysik in K\"{o}ln for support.

\end{document}